\documentclass[aps,prl,preprint,superscriptaddress]{revtex4-1}

\usepackage{amsmath,amsfonts,amssymb}
\usepackage{newtxtext,newtxmath}	

\usepackage[pdftex]{graphicx}
\usepackage{gensymb}
\usepackage{microtype}
\usepackage{bm}
\usepackage[normalem]{ulem}

\usepackage[svgnames]{xcolor}
\usepackage[
	colorlinks=True,linkcolor=MediumBlue,citecolor=MediumBlue,urlcolor=MediumBlue,
	pdfstartview=FitH,bookmarks=False,pdfpagemode=UseNone
]{hyperref}

\begin{document}

\title{\boldmath Terahertz sum-frequency excitation of a Raman-active phonon}

\author{Sebastian Maehrlein}
\email{maehrlein@fhi-berlin.mpg.de}
\affiliation{Department of Physical Chemistry, Fritz-Haber-Institut, Faradayweg 4-6, Berlin 14915, Germany}
\author{Alexander Paarmann}
\affiliation{Department of Physical Chemistry, Fritz-Haber-Institut, Faradayweg 4-6, Berlin 14915, Germany}
\author{Martin Wolf}
\affiliation{Department of Physical Chemistry, Fritz-Haber-Institut, Faradayweg 4-6, Berlin 14915, Germany}
\author{Tobias Kampfrath}
\affiliation{Department of Physical Chemistry, Fritz-Haber-Institut, Faradayweg 4-6, Berlin 14915, Germany}

\date{\today}

\begin{abstract} 
In stimulated Raman scattering, two incident optical waves induce a force oscillating at the difference of the two light frequencies. This process has enabled important applications such as the excitation and coherent control of phonons and magnons by femtosecond laser pulses. Here, we experimentally and theoretically demonstrate the so far neglected up-conversion counterpart of this process: THz sum-frequency excitation of a Raman-active phonon mode, which is tantamount to two-photon absorption by an optical transition between two adjacent vibrational levels. Coherent control of an optical lattice vibration of diamond is achieved by an intense terahertz pulse whose spectrum is centered at half the phonon frequency of 40\,THz. Remarkably, the carrier-envelope phase of the driving pulse is directly imprinted on the lattice vibration. New prospects in infrared spectroscopy, light storage schemes and lattice trajectory control in the electronic ground state emerge.
\end{abstract}

\maketitle

Gaining control over ultrafast elementary motions of electrons, spins and ions in solids is of major interest from a fundamental as well as an applied point of view. Examples include steering of chemical reactions along a desired pathway \cite{LaRue2015} and switching of spins in magnetically ordered materials\cite{Kubacka2014,Kimel2005}. Many control strategies take advantage of low-energy, that is, terahertz (THz) modes of the material. An important and ubiquitous example are lattice vibrations (phonons) whose ultrafast control by femtosecond laser pulses \cite{Cheng1990,Cho1990} has enabled new pathways to permanent material modification \cite{Klieber2011}, insulator-to-metal transitions\cite{Kim2012a,Rini2007}, magnetization manipulation \cite{Kim2012,Kubacka2014} and even memories for light \cite{England2013}. In addition, phonons are considered model systems for magnons, plasmons and other harmonic-oscillator-type modes.

\begin{figure}[tb]
\includegraphics[width=0.6\columnwidth]{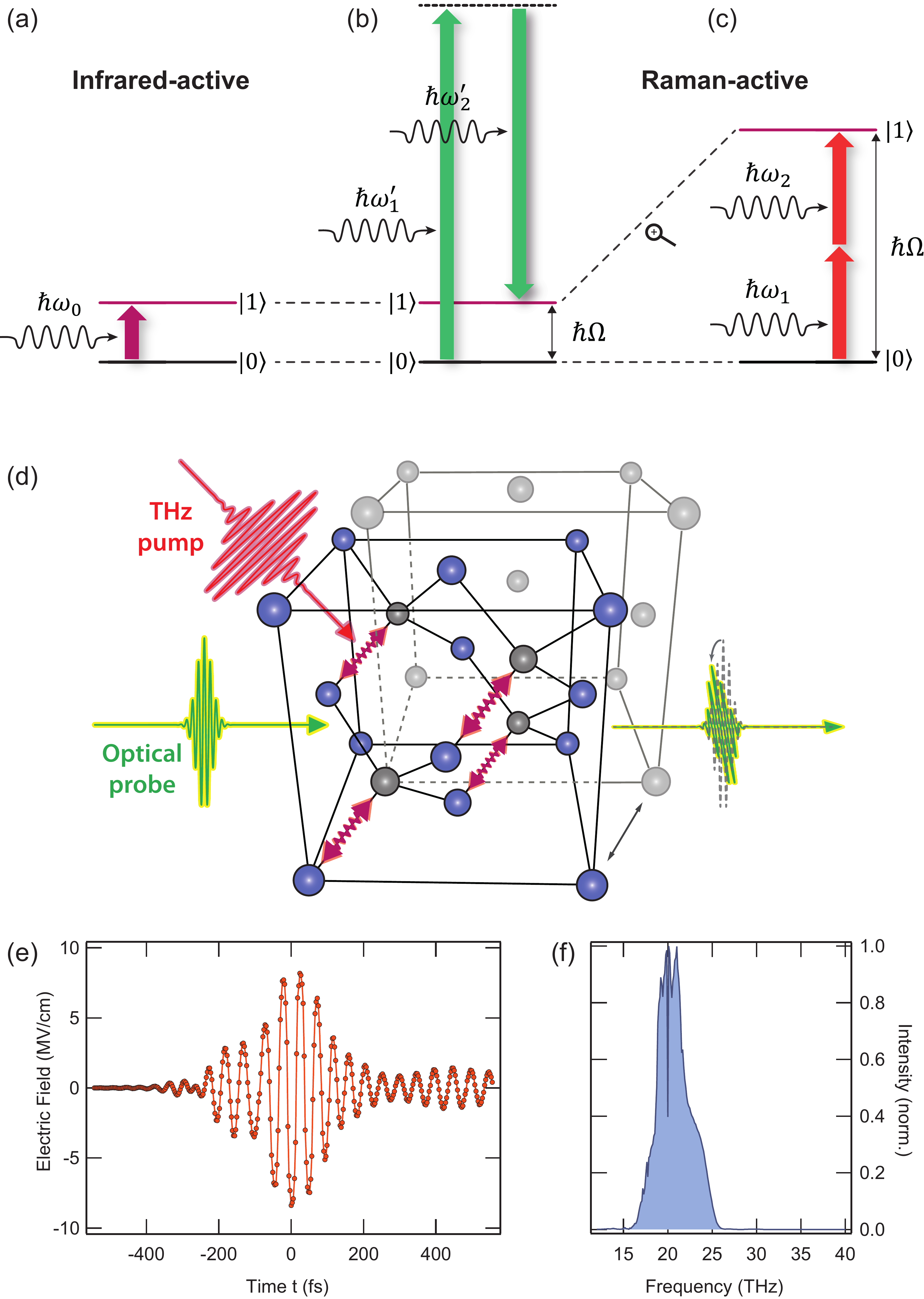}
\caption{\label{fig:Figure1}
(a)--(c) Schemes for driving coherent phonons. (a) Direct excitation: the pump frequency is resonant with the infrared-active phonon mode. (b) Stimulated Raman scattering (SRS): the difference frequency of two spectral components of the pump field is resonant with a Raman-active phonon mode. (c) Sum-frequency excitation (SFE): the sum frequency of two incident THz electric-field components is resonant with a Raman-active phonon mode. This so far neglected process can be considered as 2PA by an optical transition between two adjacent vibrational levels. (d) Unit cell of diamond and schematic of the experiment: an intense THz pulse (red) excites the high-frequency $F_{2g}$-mode whose dynamics are monitored by a delayed ultrashort optical probe pulse (green). Red arrows indicate coherent lattice motion corresponding to the Raman-active, infrared-forbidden $F_{2g}$-mode at 40\,THz. (e) Transient electric field and (f) intensity spectrum of a typical THz pump pulse used in the experiments.
}
\end{figure}

For selective excitation of coherent lattice vibrations, schemes based on one- or two-photon interactions are established [see Figs.~\ref{fig:Figure1}(a),(b)]. In the most straightforward approach, the laser electric field directly drives the electric dipoles of a long-wavelength lattice mode. This one-photon absorption process requires an infrared-active transition and a pump spectrum that covers the phonon frequency $\Omega/2\pi$ typically located between $\sim$1 and 40\,THz \cite{Huber2015,Katayama2012,Qi2009,Jewariya2010} [Fig.~\ref{fig:Figure1}(a)]. To date, many works have implemented lattice excitation using stimulated Raman scattering (SRS) of a light pulse containing electric-field components with frequencies $\omega_1'$ and $\omega_2'=\omega_1'+\Omega$, both at typically $\sim 2\pi \cdot 500$\,THz [Fig.~\ref{fig:Figure1}(b)]. The rapidly oscillating light field couples primarily to electrons of the material \cite{Merlin1997}, either nonresonantly (in optically transparent solids) \cite{Dhar1994,Shen1965} or resonantly (in opaque solids) \cite{Bothschafter2013,Dekorsy2000}, resulting in a force that coherently drives the lattice at the difference frequency $\omega_2'-\omega_1' = \Omega$. Such rectification can also be achieved through anharmonic interaction with infrared-active phonons at higher frequencies, which are excited resonantly by a THz pump pulse \cite{Foerst2011}. 

Note that the difference-frequency process underlying stimulated Raman excitation [Fig.~\ref{fig:Figure1}(b)] is a familiar phenomenon in nonlinear optics where it is also accompanied by the generation of signals at the sum of the incident frequencies. This analogy suggests a novel scheme of coherent lattice control where the two incident frequencies are chosen such that the sum frequency is resonant with the target mode [Fig.~\ref{fig:Figure1}(c)]. Such sum-frequency excitation (SFE) by a single light pulse is tantamount to two-photon absorption (2PA), which has been widely used to induce \cite{Gibson1976,Seo2011} and coherently control \cite{Meshulach1998,Silberberg2009} electronic transitions by optical photons of $\sim$1\,eV energy. So far, however, SFE or 2PA by vibrational resonances such as optical phonons have not yet been demonstrated because the laser frequencies used ($\sim$500\,THz) were typically one order of magnitude larger than the highest phonon frequencies ($\sim$40\,THz).

In this work, we provide a first experimental and theoretical demonstration of SFE of lattice vibrations. An intense THz pulse with a center frequency of 20\,THz is found to coherently drive the long-wavelength Raman-active optical phonon of diamond at the sum frequency of 40\,THz. Remarkably, we observe that the carrier-envelope phase (CEP) of the THz pulse is directly transferred into the phase of the lattice vibration. Due to the low THz photon energy, parasitic electronic excitations are suppressed, thereby opening up the way to coherent control of fundamental Raman-active modes such as phonons and magnons in the electronic ground state.

\textbf{Experiment.} To control and monitor coherent phonons by SFE, we make use of the pump-probe scheme shown in Fig.~\ref{fig:Figure1}(d). An intense phase-locked THz pulse \cite{Sell2008} with tunable center frequency is directed onto the sample to drive coherent lattice vibrations. The instantaneous time-dependent phonon amplitude is monitored by a time-delayed optical probe pulse that measures the transient birefringence of the material \cite{Cheng1990,Cho1990}. 

As sample, we choose a high-purity single crystal of diamond(100) (type IIa, thickness of 200 $\mu$m, Sumitomo Electric Hardmetal Corporation) grown by chemical vapor deposition. We focus on the zone-center optical phonon at $\Omega/2\pi \approx 40$\,THz, which is one of the highest phonon frequencies known. This $F_{2g}$ $(\Gamma_{25}^+)$-type mode is characterized by symmetric stretch vibrations of the C--C bonds [Fig.~\ref{fig:Figure1}(d)] and exhibits, therefore, no transient electric dipole moment. It is, however, Raman-active and has frequently served as a benchmark resonance in the history of Raman studies \cite{Ishioka2006,Raman1928,Solin1970}. The model character of this mode is underlined by the fact that related high-frequency vibrations are found in many other C--C covalent-bond systems such as carbon nanotubes \cite{Gambetta2006}, graphene \cite{Ishioka2008} or organic molecules. 

To drive and observe the high-frequency diamond phonon, intense phase-locked THz pulses and very short probe pulses are required. The THz pump pulse is generated by difference-frequency mixing of two phase-correlated infrared pulses from two optical parametric amplifiers each pumped by a laser pulse (duration 40\,fs, energy 7\,mJ, center wavelength 800\,nm, repetition rate 1\,kHz) from the same Ti:sapphire laser system in a GaSe crystal \cite{Sell2008} (thickness of 1\,mm). A typical THz pulse and its intensity spectrum are shown in Figs.~\ref{fig:Figure1}(e) and (f), respectively. The pulse is inherently phase-locked, and its CEP can be set by slightly varying the delay between the two generating infrared pulses \cite{Sell2008}. The pump-induced coherent-phonon dynamics are sampled by a synchronized ultrashort probe pulse that measures the transient birefringence of the sample. The probe pulse (duration 7\,fs, center wavelength 750\,nm) is derived from the seed-laser oscillator of the amplifier laser system, traverses the sample under normal incidence and collinearly with the pump, and finally enters a balanced optical bridge\cite{Dhar1994} detecting its polarization state. The linear polarizations of pump und probe are set to an angle of 45$^\circ$ with respect to each other. To determine the electric field of the THz pulse by electrooptic sampling \cite{Leitenstorfer1999,Wu1996}, the diamond sample is replaced by a thin GaSe detection crystal \cite{Sell2008} (thickness of $\sim$30\,$\mu$m), resulting in the data shown by Figs.~\ref{fig:Figure1}(e)--(f). All experiments are performed at room temperature and under ambient conditions.


\begin{figure*}[t]
\includegraphics[width=0.9\textwidth]{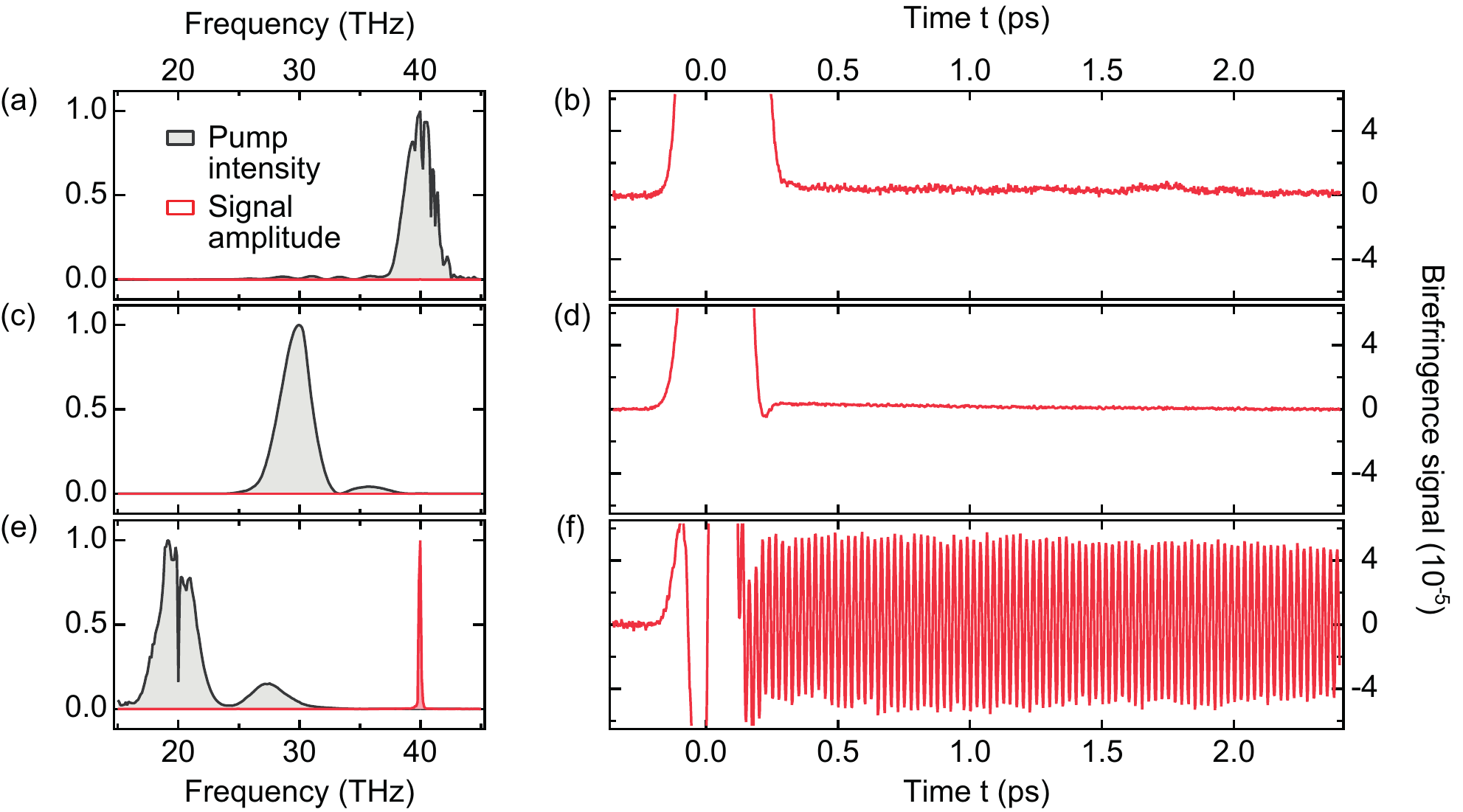}
\caption{\label{fig:Figure2}
Sum-frequency excitation of a coherent phonon. (a) Spectrum of a pump pulse centered at the $F_{2g}$-phonon frequency $\Omega/2\pi\approx$\,40\,THz. (b) Resulting pump-probe signal, showing no signature of lattice vibrations. The peak around $t=0$ is due to the instantaneous electronic response to the pump field. Panels (c), (d) and (e), (f) are as with (a), (b) but for a pump center frequency of $0.5\,\Omega$ and $0.75\,\Omega$, respectively. In panel (f), a long-lived oscillation reveals a coherent phonon at frequency $\Omega/2\pi\approx$\,40\,THz as seen in the signal spectrum of panel (e). The sharp dip in the pump spectrum of panel (e) arises from absorption by CO$_2$ in ambient air. All birefringence signals are normalized to an incident pump-pulse energy of 1\,$\mu$J.
}
\end{figure*}

\textbf{Coherent phonon signals.} Figure~\ref{fig:Figure2} shows transient birefringence signals $S(t)$ obtained for three different center frequencies $\omega_0$ of the pump pulse: $\Omega$, 0.75\,$\Omega$ and 0.5\,$\Omega$. When the pump center frequency is resonant with the target mode [Fig.~\ref{fig:Figure2}(a)], we observe a peak around $t=0$ whose shape approximately follows the intensity envelope of the pump pulse [Fig.~\ref{fig:Figure2}(b)]. This transient optical birefringence arises from the instantaneous nonresonant response of the diamond electrons to the THz pump pulse. Apart from this Kerr effect \cite{Hoffmann2009}, however, we do not observe any features that would be indicative of a coherent lattice vibration. When we lower the pump center frequency from $\omega_0=\Omega$ to 0.75\,$\Omega$ [Fig.~\ref{fig:Figure2}(c)], very similar dynamics are found [Fig.~\ref{fig:Figure2}(d)].
 
A strikingly different response is, however, obtained when the pump spectrum is centered at half the phonon frequency [Fig.~\ref{fig:Figure2}(e)]: for delays $t$ larger than the pump-pulse duration, the transient birefringence is dominated by an oscillatory component [Fig.~\ref{fig:Figure2}(f)] whose Fourier spectrum exhibits a sharp peak located at 40\,THz [Fig.~\ref{fig:Figure2}(e)]. By fitting an exponentially damped harmonic oscillation to $S(t)$ for times $t>200$\,fs, we obtain a center frequency of $\Omega/2\pi= 39.95$\,THz and an exponential damping constant of 0.283\,ps$^{-1}$ (see Supplementary Fig. S1). These values agree excellently with those inferred from SRS excitation [Fig.~\ref{fig:Figure1}(b)] using optical pump pulses with a center frequency of 750\,THz \cite{Ishioka2006}, as well as from spontaneous Raman scattering \cite{Solin1970} (see Supplementary Table S1). Therefore, the oscillatory pump-probe signal of Fig.~\ref{fig:Figure2}(f) is a clear hallmark of diamond's $F_{2g}$-phonon.
As seen in Fig. 3(a), the oscillation amplitude is found to grow linearly with the pump power, that is, the square of the driving THz field. Therefore, the coherent lattice vibration is the result of two interactions with the THz pump field and can only arise from a difference-frequency process [SRS, Fig.~\ref{fig:Figure1}(b)] or a sum-frequency process [SFE, Fig.~\ref{fig:Figure1}(c)]. Since the coherent phonon is observed for a pump center frequency of $\Omega/2$ but not $\Omega$, we have provided evidence for the sum-frequency process suggested in Fig.~\ref{fig:Figure1}(c). In other words, we here observe a coherent lattice vibration driven by SFE/2PA of a THz pump pulse. Importantly, the transient signals do not exhibit any indications of unwanted sample heating in the form of a slowly varying background, as is often observed with optical excitation \cite{Cheng1990,Cho1990,Dhar1994}.


\textbf{Model.} To put our interpretation on a theoretical basis, we assume that the solid has only one uniform vibrational mode, an optical phonon characterized by the normal coordinate $Q$ and frequency $\Omega$. The electric field $E(t)$ of the incident pump pulse exerts Coulomb forces on electrons and ions, resulting in a time-dependent electric-dipole density (polarization) $P(t)$. Since diamond is an insulator (electronic band gap of 5.5\,eV), the material remains in its electronic ground state, and the electron density follows the pump field (photon energy $\sim$\,80\ldots160\,meV) adiabatically. Consequently, the polarization is fully determined by the instantaneous values of $Q$ and $E$, and we have $P(t)=P\left(Q(t),E(t)\right)$. To describe the dynamics of the solid, we assume a Lagrangian \cite{Shen1965} that captures the mode's kinetic energy ($\dot{Q}^2/2$), potential energy ($\Omega^2 Q^2/2$) and interaction ($PE$) with the pump field. Linearization of $P$ with respect to $E$ and using the Lagrange equation finally yields the harmonic-oscillator-type equation of motion for $Q$,
	\begin{equation}\label{eq:Equation1}
		\left( \frac{\partial^2}{\partial t^2} + \Gamma \frac{\partial}{\partial t} + \Omega^2 \right) Q = E \frac{\partial P}{\partial Q} + E^2 \frac{\partial \chi}{\partial Q}.
	\end{equation}
Its right-hand side is the sum of two driving forces. The first term describes the direct coupling of $E$ to the polarization arising from lattice displacement. It corresponds to the scheme of Fig.~\ref{fig:Figure1}(a), but vanishes here because the effective charge $\partial P/\partial Q$ of the infrared-inactive $F_{2g}$-mode of diamond is zero. The second term results from the coupling of $E$ to the $Q$-dependent part of the field-induced electronic polarization $\chi E$ where $\chi = \partial P / \partial E$ is the linear dielectric susceptibility. Clearly, this coupling is Raman-type because it is mediated by the Raman tensor $\partial \chi / \partial Q$, which is nonzero for the relevant diamond phonon mode \cite{Ishioka2006,Solin1970}.

\begin{figure*}[t]
\includegraphics[width=0.9\textwidth]{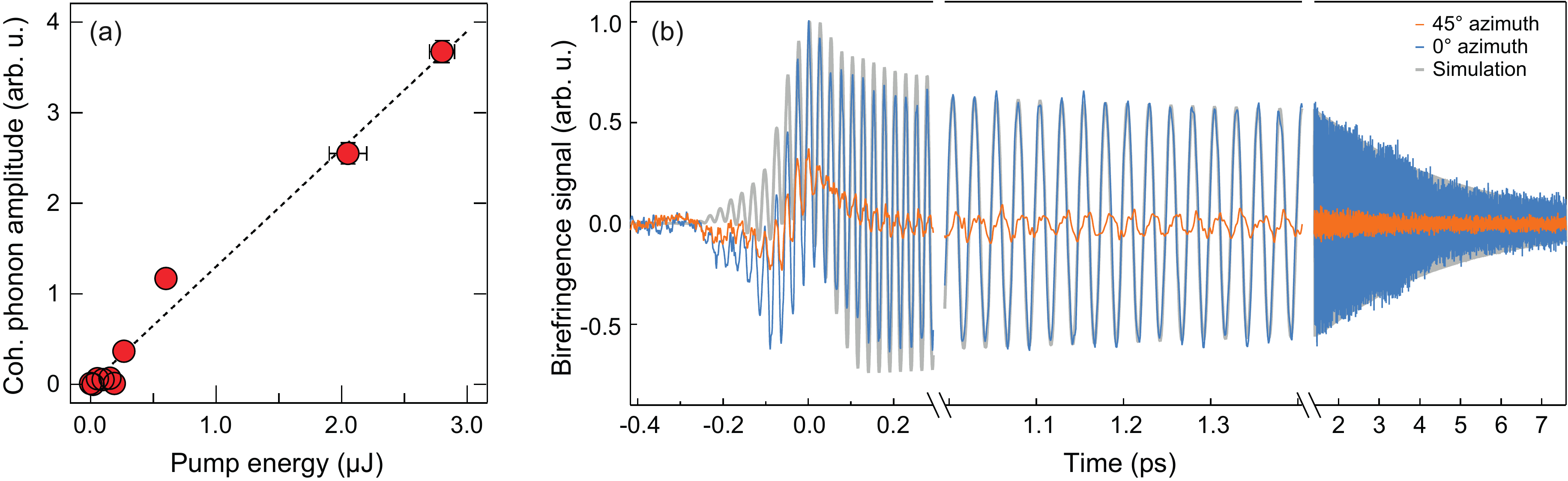}
\caption{\label{fig:Figure3}
(a) Pump-energy dependence of the coherent phonon amplitude. (b) Measured (blue line) and simulated (grey line) THz-pulse-induced phonon dynamics. The simulation is based on a harmonic oscillator model with a driving force $\propto E^2 (t)$ [see Eq.~(\ref{eq:Equation1}) and text for details]. Azimuthal sample rotation by 45$^\circ$ leads to an almost vanishing phonon signal (orange curve), consistent with the Raman-tensor symmetry of diamond. The instantaneous electronic response ($t \approx 0$) is less prominent than in Fig.~\ref{fig:Figure2} since an optimized pump spectrum [Fig.~\ref{fig:Figure1}(e)] was used.
}
\end{figure*}

Note that the second force term on the right-hand side of Eq.~(\ref{eq:Equation1}) follows the squared pump field $E^2 (t)$. Therefore, it consists of a superposition of difference- and sum-frequency pairs of the spectrum of $E(t)$. The difference-frequency component, corresponding to conventional SRS [Fig.~\ref{fig:Figure1}(b)], follows the pump intensity envelope. It has been employed for impulsive excitation of phonons \cite{Bothschafter2013,Cheng1990,Cho1990} and magnons \cite{Kimel2005} in numerous works. As we show here, also the sum-frequency component is capable of driving such Raman-active modes [Fig.~\ref{fig:Figure1}(c)]. It is precisely this term that describes phonon excitation by THz SFE. 

We solve Eq.~(\ref{eq:Equation1}) for $Q(t)$ and model the transient optical birefringence as a weighted sum of $Q(t)$ and $E^2 (t)$, the latter capturing the instantaneous electronic response (THz Kerr effect \cite{Hoffmann2009}). A final temporal convolution with a square function accounts for the velocity mismatch of pump and probe pulses propagating through the diamond sample \cite{Sajadi2015}. Using this procedure together with the actual THz driving field of Fig.~\ref{fig:Figure1}(e), we obtain the grey solid line of Fig.~\ref{fig:Figure3}(b). The agreement with the measured signal (blue solid line) is striking, in particular during temporal overlap of pump and probe pulses. 

In addition, the scaling of the pump-probe signal with the Raman tensor [see Eq.~(\ref{eq:Equation1})] is consistent with its dependence on sample orientation \cite{Ishioka2006,Solin1970}. For example, when the sample is rotated from the azimuth of largest phonon signal amplitude [blue line in Fig.~\ref{fig:Figure3}(b)] by an angle of 45$^\circ$, we find almost complete suppression of the oscillatory component [orange line in Fig.~\ref{fig:Figure3}(b)], in agreement with the symmetry of the (100)-oriented diamond crystal \cite{Ishioka2006}. Therefore, our results fully support the notion that we have observed THz SFE as a new mechanism for launching of a coherent phonon.


\begin{figure}[b]
\includegraphics[width=0.9\columnwidth]{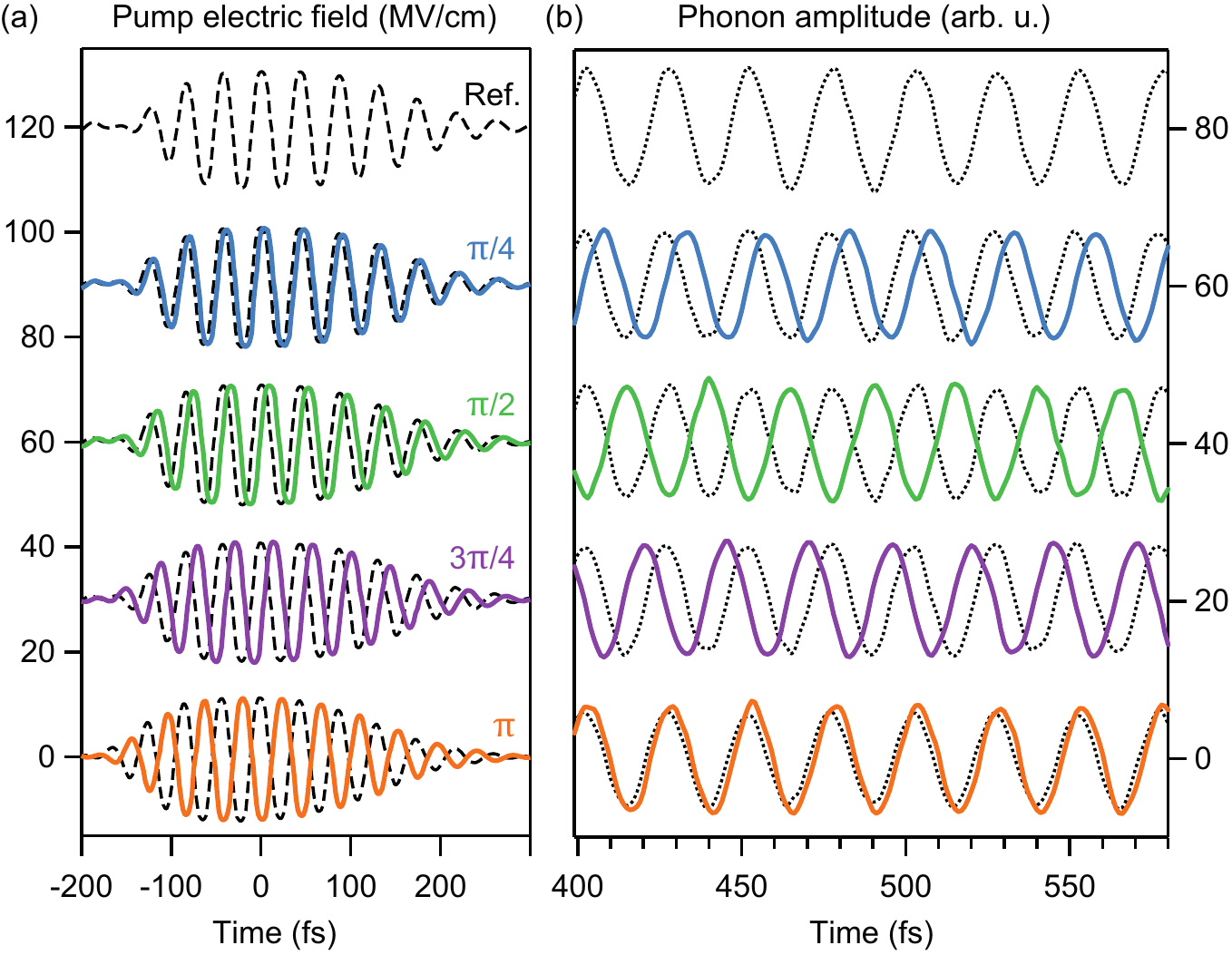}
\caption{\label{fig:Figure4}
Demonstration of phonon phase control by tuning the THz pump CEP (a) Measured transient electric field of the THz pump pulse for five CEP settings from 0 to $\pi$ (solid lines). For comparison, the $\Delta \varphi_0 = 0$ reference transient (broken lines) is measured simultaneously for each CEP. (b) Corresponding coherent phonon signals (solid lines) measured after sample excitation with the pump fields of panel (a) including corresponding reference phonon signals (broken lines).
}
\end{figure}

\textbf{Coherent-phonon phase control.} Our experiment allows us to explore the impact of the CEP on the SFE process, which is expected to be very different from the SRS case. Note that a CEP change by $\Delta \varphi_0$ shifts the phase of the phonon oscillation by $\Delta \varphi_0 - \Delta \varphi_0=0$ in the difference-frequency-type case (SRS), but by $\Delta \varphi_0 + \Delta \varphi_0= 2 \Delta \varphi_0$ in the sum-frequency process (SFE). In other words, conventional SRS [(Fig.~\ref{fig:Figure1}(b)] is independent of the CEP whereas its sum-frequency counterpart [Fig.~\ref{fig:Figure1}(c)] undergoes a shift of twice the CEP. Consequently, THz SFE should enable direct phase control of coherent phonon motion by tuning the CEP of the pump pulse.

To put this expectation to test, we measure phonon dynamics as a function of the CEP as shown in Fig.~\ref{fig:Figure4}. For example, when the CEP changes by $\Delta \varphi_0 = \pi/2$ [Fig.~\ref{fig:Figure4}(a), green line], the phonon oscillation is found to undergo a phase shift of $\pi$ [Fig.~\ref{fig:Figure4}(b), green line]. The results of Fig.~\ref{fig:Figure4} unambiguously demonstrate that twice the CEP shift of the pump field is directly transferred into the phase shift of the resulting lattice motion.

\textbf{Discussion.} Neither 2PA nor SFE have so far been observed for an optical transition between adjacent vibrational levels. Here, we witness the full dynamics of this process, starting with the coherent SFE of the vibration, the subsequent decoherence and absorption. In contrast to previous coherent quantum-control studies of 2PA by electronic transitions \cite{Meshulach1998,Silberberg2009,Zhang2009}, the excited coherent-phonon state of this work has a much lower frequency and longer dephasing times. Therefore, the coherent dynamics of the excited state can be sampled by a femtosecond optical probe directly in the time domain. 

In terms of applications, the very selective energy deposition in Raman-active modes by THz 2PA may enable machining of transparent materials by intense mid-infrared industrial lasers. Moreover, established, highly sensitive infrared vibrational spectroscopies (such as action spectroscopy \cite{Oomens2006}) can be extended to infrared-forbidden yet Raman-active modes.
Remarkably, due to the low photon energies required, THz SFE provides a unique tool to steer chemical reactions or phase transitions in the electronic ground state, solely driven by Raman-active lattice vibrations. Shaping of THz pulses \cite{Eickemeyer2000} to multicolor pulse sequences could enable independent phase and amplitude control of the lattice trajectory along multiple normal coordinates, which is challenging with optical pulses accompanied by parasitic electronic excitations \cite{Katsuki2013}. THz SFE may therefore reveal novel routes to phase transitions \cite{Kim2012a,Rini2007} or other phonon-mediated effects. Its phase sensitivity enables storing the CEP of a light pulse in a coherent phonon mode \cite{England2013}, an interesting perspective for phonon-based quantum processing at room temperature.

In conclusion, we have demonstrated that coherent lattice motion can be launched and controlled by THz SFE. This approach provides coherent control of all Raman-active modes (such as phonons, magnons and plasmons) that have so far been inaccessible by THz or mid-infrared radiation. Simultaneously, parasitic electronic excitation, which occurs when optical laser pulses are used, is greatly reduced. Finally, this so far neglected type of light-matter coupling extends vibrational spectroscopy methods to infrared-forbidden modes and paves the way for CEP-sensitive light-storage devices at room temperature and full lattice trajectory control.

\begin{acknowledgments}

\textbf{Acknowledgments.} We thank Kunie Ishioka for providing us with the diamond sample. The European Union is gratefully acknowledged for funding through the ERC H2020 CoG project TERAMAG/grant no. 681917.
\end{acknowledgments}


\begin{thebibliography}{38}%
\makeatletter
\providecommand \@ifxundefined [1]{%
 \@ifx{#1\undefined}
}%
\providecommand \@ifnum [1]{%
 \ifnum #1\expandafter \@firstoftwo
 \else \expandafter \@secondoftwo
 \fi
}%
\providecommand \@ifx [1]{%
 \ifx #1\expandafter \@firstoftwo
 \else \expandafter \@secondoftwo
 \fi
}%
\providecommand \natexlab [1]{#1}%
\providecommand \enquote  [1]{``#1''}%
\providecommand \bibnamefont  [1]{#1}%
\providecommand \bibfnamefont [1]{#1}%
\providecommand \citenamefont [1]{#1}%
\providecommand \href@noop [0]{\@secondoftwo}%
\providecommand \href [0]{\begingroup \@sanitize@url \@href}%
\providecommand \@href[1]{\@@startlink{#1}\@@href}%
\providecommand \@@href[1]{\endgroup#1\@@endlink}%
\providecommand \@sanitize@url [0]{\catcode `\\12\catcode `\$12\catcode
  `\&12\catcode `\#12\catcode `\^12\catcode `\_12\catcode `\%12\relax}%
\providecommand \@@startlink[1]{}%
\providecommand \@@endlink[0]{}%
\providecommand \url  [0]{\begingroup\@sanitize@url \@url }%
\providecommand \@url [1]{\endgroup\@href {#1}{\urlprefix }}%
\providecommand \urlprefix  [0]{URL }%
\providecommand \Eprint [0]{\href }%
\providecommand \doibase [0]{http://dx.doi.org/}%
\providecommand \selectlanguage [0]{\@gobble}%
\providecommand \bibinfo  [0]{\@secondoftwo}%
\providecommand \bibfield  [0]{\@secondoftwo}%
\providecommand \translation [1]{[#1]}%
\providecommand \BibitemOpen [0]{}%
\providecommand \bibitemStop [0]{}%
\providecommand \bibitemNoStop [0]{.\EOS\space}%
\providecommand \EOS [0]{\spacefactor3000\relax}%
\providecommand \BibitemShut  [1]{\csname bibitem#1\endcsname}%
\let\auto@bib@innerbib\@empty
\bibitem [{\citenamefont {LaRue}\ \emph {et~al.}(2015)\citenamefont {LaRue},
  \citenamefont {Katayama}, \citenamefont {Lindenberg}, \citenamefont {Fisher},
  \citenamefont {\"Ostr\"om}, \citenamefont {Nilsson},\ and\ \citenamefont
  {Ogasawara}}]{LaRue2015}%
  \BibitemOpen
  \bibfield  {author} {\bibinfo {author} {\bibfnamefont {J.~L.}\ \bibnamefont
  {LaRue}}, \bibinfo {author} {\bibfnamefont {T.}~\bibnamefont {Katayama}},
  \bibinfo {author} {\bibfnamefont {A.}~\bibnamefont {Lindenberg}}, \bibinfo
  {author} {\bibfnamefont {A.~S.}\ \bibnamefont {Fisher}}, \bibinfo {author}
  {\bibfnamefont {H.}~\bibnamefont {\"Ostr\"om}}, \bibinfo {author}
  {\bibfnamefont {A.}~\bibnamefont {Nilsson}}, \ and\ \bibinfo {author}
  {\bibfnamefont {H.}~\bibnamefont {Ogasawara}},\ }\href
  {http://link.aps.org/doi/10.1103/PhysRevLett.115.036103} {\bibfield
  {journal} {\bibinfo  {journal} {Phys. Rev. Lett.}\ }\textbf {\bibinfo
  {volume} {115}},\ \bibinfo {pages} {036103} (\bibinfo {year}
  {2015})}\BibitemShut {NoStop}%
\bibitem [{\citenamefont {Kubacka}\ \emph {et~al.}(2014)\citenamefont
  {Kubacka}, \citenamefont {Johnson}, \citenamefont {Hoffmann}, \citenamefont
  {Vicario}, \citenamefont {de~Jong}, \citenamefont {Beaud}, \citenamefont
  {Gr\"ubel}, \citenamefont {Huang}, \citenamefont {Huber}, \citenamefont
  {Patthey}, \citenamefont {Chuang}, \citenamefont {Turner}, \citenamefont
  {Dakovski}, \citenamefont {Lee}, \citenamefont {Minitti}, \citenamefont
  {Schlotter}, \citenamefont {Moore}, \citenamefont {Hauri}, \citenamefont
  {Koohpayeh}, \citenamefont {Scagnoli}, \citenamefont {Ingold}, \citenamefont
  {Johnson},\ and\ \citenamefont {Staub}}]{Kubacka2014}%
  \BibitemOpen
  \bibfield  {author} {\bibinfo {author} {\bibfnamefont {T.}~\bibnamefont
  {Kubacka}}, \bibinfo {author} {\bibfnamefont {J.~A.}\ \bibnamefont
  {Johnson}}, \bibinfo {author} {\bibfnamefont {M.~C.}\ \bibnamefont
  {Hoffmann}}, \bibinfo {author} {\bibfnamefont {C.}~\bibnamefont {Vicario}},
  \bibinfo {author} {\bibfnamefont {S.}~\bibnamefont {de~Jong}}, \bibinfo
  {author} {\bibfnamefont {P.}~\bibnamefont {Beaud}}, \bibinfo {author}
  {\bibfnamefont {S.}~\bibnamefont {Gr\"ubel}}, \bibinfo {author}
  {\bibfnamefont {S.-W.}\ \bibnamefont {Huang}}, \bibinfo {author}
  {\bibfnamefont {L.}~\bibnamefont {Huber}}, \bibinfo {author} {\bibfnamefont
  {L.}~\bibnamefont {Patthey}}, \bibinfo {author} {\bibfnamefont {Y.-D.}\
  \bibnamefont {Chuang}}, \bibinfo {author} {\bibfnamefont {J.~J.}\
  \bibnamefont {Turner}}, \bibinfo {author} {\bibfnamefont {G.~L.}\
  \bibnamefont {Dakovski}}, \bibinfo {author} {\bibfnamefont {W.-S.}\
  \bibnamefont {Lee}}, \bibinfo {author} {\bibfnamefont {M.~P.}\ \bibnamefont
  {Minitti}}, \bibinfo {author} {\bibfnamefont {W.}~\bibnamefont {Schlotter}},
  \bibinfo {author} {\bibfnamefont {R.~G.}\ \bibnamefont {Moore}}, \bibinfo
  {author} {\bibfnamefont {C.~P.}\ \bibnamefont {Hauri}}, \bibinfo {author}
  {\bibfnamefont {S.~M.}\ \bibnamefont {Koohpayeh}}, \bibinfo {author}
  {\bibfnamefont {V.}~\bibnamefont {Scagnoli}}, \bibinfo {author}
  {\bibfnamefont {G.}~\bibnamefont {Ingold}}, \bibinfo {author} {\bibfnamefont
  {S.~L.}\ \bibnamefont {Johnson}}, \ and\ \bibinfo {author} {\bibfnamefont
  {U.}~\bibnamefont {Staub}},\ }\href
  {http://science.sciencemag.org/content/343/6177/1333.abstract} {\bibfield
  {journal} {\bibinfo  {journal} {Science}\ ,\ \bibinfo {pages} {1242862}}
  (\bibinfo {year} {2014})}\BibitemShut {NoStop}%
\bibitem [{\citenamefont {Kimel}\ \emph {et~al.}(2005)\citenamefont {Kimel},
  \citenamefont {Kirilyuk}, \citenamefont {Usachev}, \citenamefont {Pisarev},
  \citenamefont {Balbashov},\ and\ \citenamefont {Rasing}}]{Kimel2005}%
  \BibitemOpen
  \bibfield  {author} {\bibinfo {author} {\bibfnamefont {A.~V.}\ \bibnamefont
  {Kimel}}, \bibinfo {author} {\bibfnamefont {A.}~\bibnamefont {Kirilyuk}},
  \bibinfo {author} {\bibfnamefont {P.~A.}\ \bibnamefont {Usachev}}, \bibinfo
  {author} {\bibfnamefont {R.~V.}\ \bibnamefont {Pisarev}}, \bibinfo {author}
  {\bibfnamefont {A.~M.}\ \bibnamefont {Balbashov}}, \ and\ \bibinfo {author}
  {\bibfnamefont {T.}~\bibnamefont {Rasing}},\ }\href
  {http://dx.doi.org/10.1038/nature03564} {\bibfield  {journal} {\bibinfo
  {journal} {Nature}\ }\textbf {\bibinfo {volume} {435}},\ \bibinfo {pages}
  {655} (\bibinfo {year} {2005})}\BibitemShut {NoStop}%
\bibitem [{\citenamefont {Cheng}\ \emph {et~al.}(1990)\citenamefont {Cheng},
  \citenamefont {Brorson}, \citenamefont {Kazeroonian}, \citenamefont
  {Moodera}, \citenamefont {Dresselhaus}, \citenamefont {Dresselhaus},\ and\
  \citenamefont {Ippen}}]{Cheng1990}%
  \BibitemOpen
  \bibfield  {author} {\bibinfo {author} {\bibfnamefont {T.~K.}\ \bibnamefont
  {Cheng}}, \bibinfo {author} {\bibfnamefont {S.~D.}\ \bibnamefont {Brorson}},
  \bibinfo {author} {\bibfnamefont {A.~S.}\ \bibnamefont {Kazeroonian}},
  \bibinfo {author} {\bibfnamefont {J.~S.}\ \bibnamefont {Moodera}}, \bibinfo
  {author} {\bibfnamefont {G.}~\bibnamefont {Dresselhaus}}, \bibinfo {author}
  {\bibfnamefont {M.~S.}\ \bibnamefont {Dresselhaus}}, \ and\ \bibinfo {author}
  {\bibfnamefont {E.~P.}\ \bibnamefont {Ippen}},\ }\href {\doibase
  http://dx.doi.org/10.1063/1.104090} {\bibfield  {journal} {\bibinfo
  {journal} {Applied Physics Letters}\ }\textbf {\bibinfo {volume} {57}},\
  \bibinfo {pages} {1004} (\bibinfo {year} {1990})}\BibitemShut {NoStop}%
\bibitem [{\citenamefont {Cho}\ \emph {et~al.}(1990)\citenamefont {Cho},
  \citenamefont {K\"utt},\ and\ \citenamefont {Kurz}}]{Cho1990}%
  \BibitemOpen
  \bibfield  {author} {\bibinfo {author} {\bibfnamefont {G.~C.}\ \bibnamefont
  {Cho}}, \bibinfo {author} {\bibfnamefont {W.}~\bibnamefont {K\"utt}}, \ and\
  \bibinfo {author} {\bibfnamefont {H.}~\bibnamefont {Kurz}},\ }\href
  {http://link.aps.org/doi/10.1103/PhysRevLett.65.764} {\bibfield  {journal}
  {\bibinfo  {journal} {Phys. Rev. Lett.}\ }\textbf {\bibinfo {volume} {65}},\
  \bibinfo {pages} {764} (\bibinfo {year} {1990})}\BibitemShut {NoStop}%
\bibitem [{\citenamefont {Klieber}\ \emph {et~al.}(2011)\citenamefont
  {Klieber}, \citenamefont {Peronne}, \citenamefont {Katayama}, \citenamefont
  {Choi}, \citenamefont {Yamaguchi}, \citenamefont {Pezeril},\ and\
  \citenamefont {Nelson}}]{Klieber2011}%
  \BibitemOpen
  \bibfield  {author} {\bibinfo {author} {\bibfnamefont {C.}~\bibnamefont
  {Klieber}}, \bibinfo {author} {\bibfnamefont {E.}~\bibnamefont {Peronne}},
  \bibinfo {author} {\bibfnamefont {K.}~\bibnamefont {Katayama}}, \bibinfo
  {author} {\bibfnamefont {J.}~\bibnamefont {Choi}}, \bibinfo {author}
  {\bibfnamefont {M.}~\bibnamefont {Yamaguchi}}, \bibinfo {author}
  {\bibfnamefont {T.}~\bibnamefont {Pezeril}}, \ and\ \bibinfo {author}
  {\bibfnamefont {K.~A.}\ \bibnamefont {Nelson}},\ }\href {\doibase
  http://dx.doi.org/10.1063/1.3595275} {\bibfield  {journal} {\bibinfo
  {journal} {Applied Physics Letters}\ }\textbf {\bibinfo {volume} {98}},\
  \bibinfo {pages} {211908} (\bibinfo {year} {2011})}\BibitemShut {NoStop}%
\bibitem [{\citenamefont {Kim}\ \emph {et~al.}(2012{\natexlab{a}})\citenamefont
  {Kim}, \citenamefont {Pashkin}, \citenamefont {Sch\"afer}, \citenamefont
  {Beyer}, \citenamefont {Porer}, \citenamefont {Wolf}, \citenamefont
  {Bernhard}, \citenamefont {Demsar}, \citenamefont {Huber},\ and\
  \citenamefont {Leitenstorfer}}]{Kim2012a}%
  \BibitemOpen
  \bibfield  {author} {\bibinfo {author} {\bibfnamefont {K.~W.}\ \bibnamefont
  {Kim}}, \bibinfo {author} {\bibfnamefont {A.}~\bibnamefont {Pashkin}},
  \bibinfo {author} {\bibfnamefont {H.}~\bibnamefont {Sch\"afer}}, \bibinfo
  {author} {\bibfnamefont {M.}~\bibnamefont {Beyer}}, \bibinfo {author}
  {\bibfnamefont {M.}~\bibnamefont {Porer}}, \bibinfo {author} {\bibfnamefont
  {T.}~\bibnamefont {Wolf}}, \bibinfo {author} {\bibfnamefont {C.}~\bibnamefont
  {Bernhard}}, \bibinfo {author} {\bibfnamefont {J.}~\bibnamefont {Demsar}},
  \bibinfo {author} {\bibfnamefont {R.}~\bibnamefont {Huber}}, \ and\ \bibinfo
  {author} {\bibfnamefont {A.}~\bibnamefont {Leitenstorfer}},\ }\href
  {http://dx.doi.org/10.1038/nmat3294} {\bibfield  {journal} {\bibinfo
  {journal} {Nat Mater}\ }\textbf {\bibinfo {volume} {11}},\ \bibinfo {pages}
  {497} (\bibinfo {year} {2012}{\natexlab{a}})}\BibitemShut {NoStop}%
\bibitem [{\citenamefont {Rini}\ \emph {et~al.}(2007)\citenamefont {Rini},
  \citenamefont {Tobey}, \citenamefont {Dean}, \citenamefont {Itatani},
  \citenamefont {Tomioka}, \citenamefont {Tokura}, \citenamefont {Schoenlein},\
  and\ \citenamefont {Cavalleri}}]{Rini2007}%
  \BibitemOpen
  \bibfield  {author} {\bibinfo {author} {\bibfnamefont {M.}~\bibnamefont
  {Rini}}, \bibinfo {author} {\bibfnamefont {R.}~\bibnamefont {Tobey}},
  \bibinfo {author} {\bibfnamefont {N.}~\bibnamefont {Dean}}, \bibinfo {author}
  {\bibfnamefont {J.}~\bibnamefont {Itatani}}, \bibinfo {author} {\bibfnamefont
  {Y.}~\bibnamefont {Tomioka}}, \bibinfo {author} {\bibfnamefont
  {Y.}~\bibnamefont {Tokura}}, \bibinfo {author} {\bibfnamefont {R.~W.}\
  \bibnamefont {Schoenlein}}, \ and\ \bibinfo {author} {\bibfnamefont
  {A.}~\bibnamefont {Cavalleri}},\ }\href {\doibase 10.1038/nature06119}
  {\bibfield  {journal} {\bibinfo  {journal} {Nature}\ }\textbf {\bibinfo
  {volume} {449}},\ \bibinfo {pages} {72} (\bibinfo {year} {2007})}\BibitemShut
  {NoStop}%
\bibitem [{\citenamefont {Kim}\ \emph {et~al.}(2012{\natexlab{b}})\citenamefont
  {Kim}, \citenamefont {Vomir},\ and\ \citenamefont {Bigot}}]{Kim2012}%
  \BibitemOpen
  \bibfield  {author} {\bibinfo {author} {\bibfnamefont {J.-W.}\ \bibnamefont
  {Kim}}, \bibinfo {author} {\bibfnamefont {M.}~\bibnamefont {Vomir}}, \ and\
  \bibinfo {author} {\bibfnamefont {J.-Y.}\ \bibnamefont {Bigot}},\ }\href
  {http://link.aps.org/doi/10.1103/PhysRevLett.109.166601} {\bibfield
  {journal} {\bibinfo  {journal} {Phys. Rev. Lett.}\ }\textbf {\bibinfo
  {volume} {109}},\ \bibinfo {pages} {166601} (\bibinfo {year}
  {2012}{\natexlab{b}})}\BibitemShut {NoStop}%
\bibitem [{\citenamefont {England}\ \emph {et~al.}(2013)\citenamefont
  {England}, \citenamefont {Bustard}, \citenamefont {Nunn}, \citenamefont
  {Lausten},\ and\ \citenamefont {Sussman}}]{England2013}%
  \BibitemOpen
  \bibfield  {author} {\bibinfo {author} {\bibfnamefont {D.~G.}\ \bibnamefont
  {England}}, \bibinfo {author} {\bibfnamefont {P.~J.}\ \bibnamefont
  {Bustard}}, \bibinfo {author} {\bibfnamefont {J.}~\bibnamefont {Nunn}},
  \bibinfo {author} {\bibfnamefont {R.}~\bibnamefont {Lausten}}, \ and\
  \bibinfo {author} {\bibfnamefont {B.~J.}\ \bibnamefont {Sussman}},\ }\href
  {http://link.aps.org/doi/10.1103/PhysRevLett.111.243601} {\bibfield
  {journal} {\bibinfo  {journal} {Phys. Rev. Lett.}\ }\textbf {\bibinfo
  {volume} {111}},\ \bibinfo {pages} {243601} (\bibinfo {year}
  {2013})}\BibitemShut {NoStop}%
\bibitem [{\citenamefont {Huber}\ \emph {et~al.}(2015)\citenamefont {Huber},
  \citenamefont {Ranke}, \citenamefont {Ferrer}, \citenamefont {Huber},\ and\
  \citenamefont {Johnson}}]{Huber2015}%
  \BibitemOpen
  \bibfield  {author} {\bibinfo {author} {\bibfnamefont {T.}~\bibnamefont
  {Huber}}, \bibinfo {author} {\bibfnamefont {M.}~\bibnamefont {Ranke}},
  \bibinfo {author} {\bibfnamefont {A.}~\bibnamefont {Ferrer}}, \bibinfo
  {author} {\bibfnamefont {L.}~\bibnamefont {Huber}}, \ and\ \bibinfo {author}
  {\bibfnamefont {S.~L.}\ \bibnamefont {Johnson}},\ }\href {\doibase
  http://dx.doi.org/10.1063/1.4930021} {\bibfield  {journal} {\bibinfo
  {journal} {Applied Physics Letters}\ }\textbf {\bibinfo {volume} {107}},\
  \bibinfo {pages} {091107} (\bibinfo {year} {2015})}\BibitemShut {NoStop}%
\bibitem [{\citenamefont {Katayama}\ \emph {et~al.}(2012)\citenamefont
  {Katayama}, \citenamefont {Aoki}, \citenamefont {Takeda}, \citenamefont
  {Shimosato}, \citenamefont {Ashida}, \citenamefont {Kinjo}, \citenamefont
  {Kawayama}, \citenamefont {Tonouchi}, \citenamefont {Nagai},\ and\
  \citenamefont {Tanaka}}]{Katayama2012}%
  \BibitemOpen
  \bibfield  {author} {\bibinfo {author} {\bibfnamefont {I.}~\bibnamefont
  {Katayama}}, \bibinfo {author} {\bibfnamefont {H.}~\bibnamefont {Aoki}},
  \bibinfo {author} {\bibfnamefont {J.}~\bibnamefont {Takeda}}, \bibinfo
  {author} {\bibfnamefont {H.}~\bibnamefont {Shimosato}}, \bibinfo {author}
  {\bibfnamefont {M.}~\bibnamefont {Ashida}}, \bibinfo {author} {\bibfnamefont
  {R.}~\bibnamefont {Kinjo}}, \bibinfo {author} {\bibfnamefont
  {I.}~\bibnamefont {Kawayama}}, \bibinfo {author} {\bibfnamefont
  {M.}~\bibnamefont {Tonouchi}}, \bibinfo {author} {\bibfnamefont
  {M.}~\bibnamefont {Nagai}}, \ and\ \bibinfo {author} {\bibfnamefont
  {K.}~\bibnamefont {Tanaka}},\ }\href
  {http://link.aps.org/doi/10.1103/PhysRevLett.108.097401} {\bibfield
  {journal} {\bibinfo  {journal} {Phys. Rev. Lett.}\ }\textbf {\bibinfo
  {volume} {108}},\ \bibinfo {pages} {097401} (\bibinfo {year}
  {2012})}\BibitemShut {NoStop}%
\bibitem [{\citenamefont {Qi}\ \emph {et~al.}(2009)\citenamefont {Qi},
  \citenamefont {Shin}, \citenamefont {Yeh}, \citenamefont {Nelson},\ and\
  \citenamefont {Rappe}}]{Qi2009}%
  \BibitemOpen
  \bibfield  {author} {\bibinfo {author} {\bibfnamefont {T.}~\bibnamefont
  {Qi}}, \bibinfo {author} {\bibfnamefont {Y.-H.}\ \bibnamefont {Shin}},
  \bibinfo {author} {\bibfnamefont {K.-L.}\ \bibnamefont {Yeh}}, \bibinfo
  {author} {\bibfnamefont {K.~A.}\ \bibnamefont {Nelson}}, \ and\ \bibinfo
  {author} {\bibfnamefont {A.~M.}\ \bibnamefont {Rappe}},\ }\href
  {http://link.aps.org/doi/10.1103/PhysRevLett.102.247603} {\bibfield
  {journal} {\bibinfo  {journal} {Phys. Rev. Lett.}\ }\textbf {\bibinfo
  {volume} {102}},\ \bibinfo {pages} {247603} (\bibinfo {year}
  {2009})}\BibitemShut {NoStop}%
\bibitem [{\citenamefont {Jewariya}\ \emph {et~al.}(2010)\citenamefont
  {Jewariya}, \citenamefont {Nagai},\ and\ \citenamefont
  {Tanaka}}]{Jewariya2010}%
  \BibitemOpen
  \bibfield  {author} {\bibinfo {author} {\bibfnamefont {M.}~\bibnamefont
  {Jewariya}}, \bibinfo {author} {\bibfnamefont {M.}~\bibnamefont {Nagai}}, \
  and\ \bibinfo {author} {\bibfnamefont {K.}~\bibnamefont {Tanaka}},\ }\href
  {http://link.aps.org/doi/10.1103/PhysRevLett.105.203003} {\bibfield
  {journal} {\bibinfo  {journal} {Phys. Rev. Lett.}\ }\textbf {\bibinfo
  {volume} {105}},\ \bibinfo {pages} {203003} (\bibinfo {year}
  {2010})}\BibitemShut {NoStop}%
\bibitem [{\citenamefont {Merlin}(1997)}]{Merlin1997}%
  \BibitemOpen
  \bibfield  {author} {\bibinfo {author} {\bibfnamefont {R.}~\bibnamefont
  {Merlin}},\ }\bibfield  {booktitle} {\emph {\bibinfo {booktitle} {Highlights
  in Condensed Matter Physics and Materials Science}},\ }\href
  {http://www.sciencedirect.com/science/article/pii/S0038109896007211}
  {\bibfield  {journal} {\bibinfo  {journal} {Solid State Communications}\
  }\textbf {\bibinfo {volume} {102}},\ \bibinfo {pages} {207} (\bibinfo {year}
  {1997})}\BibitemShut {NoStop}%
\bibitem [{\citenamefont {Dhar}\ \emph {et~al.}(1994)\citenamefont {Dhar},
  \citenamefont {Rogers},\ and\ \citenamefont {Nelson}}]{Dhar1994}%
  \BibitemOpen
  \bibfield  {author} {\bibinfo {author} {\bibfnamefont {L.}~\bibnamefont
  {Dhar}}, \bibinfo {author} {\bibfnamefont {J.~A.}\ \bibnamefont {Rogers}}, \
  and\ \bibinfo {author} {\bibfnamefont {K.~A.}\ \bibnamefont {Nelson}},\
  }\bibfield  {booktitle} {\emph {\bibinfo {booktitle} {Chemical Reviews}},\
  }\href {\doibase 10.1021/cr00025a006} {\bibfield  {journal} {\bibinfo
  {journal} {Chem. Rev.}\ }\textbf {\bibinfo {volume} {94}},\ \bibinfo {pages}
  {157} (\bibinfo {year} {1994})}\BibitemShut {NoStop}%
\bibitem [{\citenamefont {Shen}\ and\ \citenamefont
  {Bloembergen}(1965)}]{Shen1965}%
  \BibitemOpen
  \bibfield  {author} {\bibinfo {author} {\bibfnamefont {Y.~R.}\ \bibnamefont
  {Shen}}\ and\ \bibinfo {author} {\bibfnamefont {N.}~\bibnamefont
  {Bloembergen}},\ }\href {http://link.aps.org/doi/10.1103/PhysRev.137.A1787}
  {\bibfield  {journal} {\bibinfo  {journal} {Phys. Rev.}\ }\textbf {\bibinfo
  {volume} {137}},\ \bibinfo {pages} {A1787} (\bibinfo {year}
  {1965})}\BibitemShut {NoStop}%
\bibitem [{\citenamefont {Bothschafter}\ \emph {et~al.}(2013)\citenamefont
  {Bothschafter}, \citenamefont {Paarmann}, \citenamefont {Zijlstra},
  \citenamefont {Karpowicz}, \citenamefont {Garcia}, \citenamefont
  {Kienberger},\ and\ \citenamefont {Ernstorfer}}]{Bothschafter2013}%
  \BibitemOpen
  \bibfield  {author} {\bibinfo {author} {\bibfnamefont {E.~M.}\ \bibnamefont
  {Bothschafter}}, \bibinfo {author} {\bibfnamefont {A.}~\bibnamefont
  {Paarmann}}, \bibinfo {author} {\bibfnamefont {E.~S.}\ \bibnamefont
  {Zijlstra}}, \bibinfo {author} {\bibfnamefont {N.}~\bibnamefont {Karpowicz}},
  \bibinfo {author} {\bibfnamefont {M.~E.}\ \bibnamefont {Garcia}}, \bibinfo
  {author} {\bibfnamefont {R.}~\bibnamefont {Kienberger}}, \ and\ \bibinfo
  {author} {\bibfnamefont {R.}~\bibnamefont {Ernstorfer}},\ }\href
  {http://link.aps.org/doi/10.1103/PhysRevLett.110.067402} {\bibfield
  {journal} {\bibinfo  {journal} {Phys. Rev. Lett.}\ }\textbf {\bibinfo
  {volume} {110}},\ \bibinfo {pages} {067402} (\bibinfo {year}
  {2013})}\BibitemShut {NoStop}%
\bibitem [{\citenamefont {Dekorsy}\ \emph {et~al.}(2000)\citenamefont
  {Dekorsy}, \citenamefont {Cho},\ and\ \citenamefont {Kurz}}]{Dekorsy2000}%
  \BibitemOpen
  \bibfield  {author} {\bibinfo {author} {\bibfnamefont {T.}~\bibnamefont
  {Dekorsy}}, \bibinfo {author} {\bibfnamefont {G.~C.}\ \bibnamefont {Cho}}, \
  and\ \bibinfo {author} {\bibfnamefont {H.}~\bibnamefont {Kurz}},\ }in\ \href
  {http://dx.doi.org/10.1007/BFb0084242} {\emph {\bibinfo {booktitle} {Topics
  in Applied Physics}}},\ Vol.~\bibinfo {volume} {76},\ \bibinfo {editor}
  {edited by\ \bibinfo {editor} {\bibfnamefont {M.}~\bibnamefont {Cardona}}\
  and\ \bibinfo {editor} {\bibfnamefont {G.}~\bibnamefont {G\"untherodt}}}\
  (\bibinfo  {publisher} {Springer Berlin Heidelberg},\ \bibinfo {year}
  {2000})\ pp.\ \bibinfo {pages} {169--209}\BibitemShut {NoStop}%
\bibitem [{\citenamefont {Foerst}\ \emph {et~al.}(2011)\citenamefont {Foerst},
  \citenamefont {Manzoni}, \citenamefont {Kaiser}, \citenamefont {Tomioka},
  \citenamefont {Tokura}, \citenamefont {Merlin},\ and\ \citenamefont
  {Cavalleri}}]{Foerst2011}%
  \BibitemOpen
  \bibfield  {author} {\bibinfo {author} {\bibfnamefont {M.}~\bibnamefont
  {Foerst}}, \bibinfo {author} {\bibfnamefont {C.}~\bibnamefont {Manzoni}},
  \bibinfo {author} {\bibfnamefont {S.}~\bibnamefont {Kaiser}}, \bibinfo
  {author} {\bibfnamefont {Y.}~\bibnamefont {Tomioka}}, \bibinfo {author}
  {\bibfnamefont {Y.}~\bibnamefont {Tokura}}, \bibinfo {author} {\bibfnamefont
  {R.}~\bibnamefont {Merlin}}, \ and\ \bibinfo {author} {\bibfnamefont
  {A.}~\bibnamefont {Cavalleri}},\ }\href {\doibase 10.1038/nphys2055}
  {\bibfield  {journal} {\bibinfo  {journal} {Nat Phys}\ }\textbf {\bibinfo
  {volume} {7}},\ \bibinfo {pages} {854} (\bibinfo {year} {2011})}\BibitemShut
  {NoStop}%
\bibitem [{\citenamefont {Gibson}\ \emph {et~al.}(1976)\citenamefont {Gibson},
  \citenamefont {Hatch}, \citenamefont {Maggs}, \citenamefont {Tilley},\ and\
  \citenamefont {Walker}}]{Gibson1976}%
  \BibitemOpen
  \bibfield  {author} {\bibinfo {author} {\bibfnamefont {A.~F.}\ \bibnamefont
  {Gibson}}, \bibinfo {author} {\bibfnamefont {C.~B.}\ \bibnamefont {Hatch}},
  \bibinfo {author} {\bibfnamefont {P.~N.~D.}\ \bibnamefont {Maggs}}, \bibinfo
  {author} {\bibfnamefont {D.~R.}\ \bibnamefont {Tilley}}, \ and\ \bibinfo
  {author} {\bibfnamefont {A.~C.}\ \bibnamefont {Walker}},\ }\href
  {http://stacks.iop.org/0022-3719/9/i=17/a=019} {\bibfield  {journal}
  {\bibinfo  {journal} {Journal of Physics C: Solid State Physics}\ }\textbf
  {\bibinfo {volume} {9}},\ \bibinfo {pages} {3259} (\bibinfo {year}
  {1976})}\BibitemShut {NoStop}%
\bibitem [{\citenamefont {Seo}\ \emph {et~al.}(2011)\citenamefont {Seo},
  \citenamefont {Gregory}, \citenamefont {Feldman}, \citenamefont {Tolk},\ and\
  \citenamefont {Cohen}}]{Seo2011}%
  \BibitemOpen
  \bibfield  {author} {\bibinfo {author} {\bibfnamefont {D.}~\bibnamefont
  {Seo}}, \bibinfo {author} {\bibfnamefont {J.~M.}\ \bibnamefont {Gregory}},
  \bibinfo {author} {\bibfnamefont {L.~C.}\ \bibnamefont {Feldman}}, \bibinfo
  {author} {\bibfnamefont {N.~H.}\ \bibnamefont {Tolk}}, \ and\ \bibinfo
  {author} {\bibfnamefont {P.~I.}\ \bibnamefont {Cohen}},\ }\href
  {http://link.aps.org/doi/10.1103/PhysRevB.83.195203} {\bibfield  {journal}
  {\bibinfo  {journal} {Phys. Rev. B}\ }\textbf {\bibinfo {volume} {83}},\
  \bibinfo {pages} {195203} (\bibinfo {year} {2011})}\BibitemShut {NoStop}%
\bibitem [{\citenamefont {Meshulach}\ and\ \citenamefont
  {Silberberg}(1998)}]{Meshulach1998}%
  \BibitemOpen
  \bibfield  {author} {\bibinfo {author} {\bibfnamefont {D.}~\bibnamefont
  {Meshulach}}\ and\ \bibinfo {author} {\bibfnamefont {Y.}~\bibnamefont
  {Silberberg}},\ }\href {http://dx.doi.org/10.1038/24329} {\bibfield
  {journal} {\bibinfo  {journal} {Nature}\ }\textbf {\bibinfo {volume} {396}},\
  \bibinfo {pages} {239} (\bibinfo {year} {1998})}\BibitemShut {NoStop}%
\bibitem [{\citenamefont {Silberberg}(2009)}]{Silberberg2009}%
  \BibitemOpen
  \bibfield  {author} {\bibinfo {author} {\bibfnamefont {Y.}~\bibnamefont
  {Silberberg}},\ }\href {\doibase 10.1146/annurev.physchem.040808.090427}
  {\bibfield  {journal} {\bibinfo  {journal} {Annu. Rev. Phys. Chem.}\ }\textbf
  {\bibinfo {volume} {60}},\ \bibinfo {pages} {277} (\bibinfo {year}
  {2009})}\BibitemShut {NoStop}%
\bibitem [{\citenamefont {Sell}\ \emph {et~al.}(2008)\citenamefont {Sell},
  \citenamefont {Leitenstorfer},\ and\ \citenamefont {Huber}}]{Sell2008}%
  \BibitemOpen
  \bibfield  {author} {\bibinfo {author} {\bibfnamefont {A.}~\bibnamefont
  {Sell}}, \bibinfo {author} {\bibfnamefont {A.}~\bibnamefont {Leitenstorfer}},
  \ and\ \bibinfo {author} {\bibfnamefont {R.}~\bibnamefont {Huber}},\ }\href
  {http://ol.osa.org/abstract.cfm?URI=ol-33-23-2767} {\bibfield  {journal}
  {\bibinfo  {journal} {Opt. Lett.}\ }\textbf {\bibinfo {volume} {33}},\
  \bibinfo {pages} {2767} (\bibinfo {year} {2008})}\BibitemShut {NoStop}%
\bibitem [{\citenamefont {Ishioka}\ \emph {et~al.}(2006)\citenamefont
  {Ishioka}, \citenamefont {Hase}, \citenamefont {Kitajima},\ and\
  \citenamefont {Petek}}]{Ishioka2006}%
  \BibitemOpen
  \bibfield  {author} {\bibinfo {author} {\bibfnamefont {K.}~\bibnamefont
  {Ishioka}}, \bibinfo {author} {\bibfnamefont {M.}~\bibnamefont {Hase}},
  \bibinfo {author} {\bibfnamefont {M.}~\bibnamefont {Kitajima}}, \ and\
  \bibinfo {author} {\bibfnamefont {H.}~\bibnamefont {Petek}},\ }\href
  {http://dx.doi.org/10.1063/1.2402231} {\bibfield  {journal} {\bibinfo
  {journal} {Appl. Phys. Lett.}\ }\textbf {\bibinfo {volume} {89}},\ \bibinfo
  {pages} {231916} (\bibinfo {year} {2006})}\BibitemShut {NoStop}%
\bibitem [{\citenamefont {Raman}\ and\ \citenamefont
  {Krishnan}(1928)}]{Raman1928}%
  \BibitemOpen
  \bibfield  {author} {\bibinfo {author} {\bibfnamefont {V.}~\bibnamefont
  {Raman}, \bibfnamefont {C.}}\ and\ \bibinfo {author} {\bibfnamefont
  {S.}~\bibnamefont {Krishnan}, \bibfnamefont {K.}},\ }\bibfield  {booktitle}
  {\emph {\bibinfo {booktitle} {Nature}},\ }\href {\doibase 10.1038/121501c0}
  {\ \textbf {\bibinfo {volume} {121}},\ \bibinfo {pages} {501} (\bibinfo
  {year} {1928})}\BibitemShut {NoStop}%
\bibitem [{\citenamefont {Solin}\ and\ \citenamefont
  {Ramdas}(1970)}]{Solin1970}%
  \BibitemOpen
  \bibfield  {author} {\bibinfo {author} {\bibfnamefont {S.~A.}\ \bibnamefont
  {Solin}}\ and\ \bibinfo {author} {\bibfnamefont {A.~K.}\ \bibnamefont
  {Ramdas}},\ }\href {http://link.aps.org/doi/10.1103/PhysRevB.1.1687}
  {\bibfield  {journal} {\bibinfo  {journal} {Phys. Rev. B}\ }\textbf {\bibinfo
  {volume} {1}},\ \bibinfo {pages} {1687} (\bibinfo {year} {1970})}\BibitemShut
  {NoStop}%
\bibitem [{\citenamefont {Gambetta}\ \emph {et~al.}(2006)\citenamefont
  {Gambetta}, \citenamefont {Manzoni}, \citenamefont {Menna}, \citenamefont
  {Meneghetti}, \citenamefont {Cerullo}, \citenamefont {Lanzani}, \citenamefont
  {Tretiak}, \citenamefont {Piryatinski}, \citenamefont {Saxena}, \citenamefont
  {Martin},\ and\ \citenamefont {Bishop}}]{Gambetta2006}%
  \BibitemOpen
  \bibfield  {author} {\bibinfo {author} {\bibfnamefont {A.}~\bibnamefont
  {Gambetta}}, \bibinfo {author} {\bibfnamefont {C.}~\bibnamefont {Manzoni}},
  \bibinfo {author} {\bibfnamefont {E.}~\bibnamefont {Menna}}, \bibinfo
  {author} {\bibfnamefont {M.}~\bibnamefont {Meneghetti}}, \bibinfo {author}
  {\bibfnamefont {G.}~\bibnamefont {Cerullo}}, \bibinfo {author} {\bibfnamefont
  {G.}~\bibnamefont {Lanzani}}, \bibinfo {author} {\bibfnamefont
  {S.}~\bibnamefont {Tretiak}}, \bibinfo {author} {\bibfnamefont
  {A.}~\bibnamefont {Piryatinski}}, \bibinfo {author} {\bibfnamefont
  {A.}~\bibnamefont {Saxena}}, \bibinfo {author} {\bibfnamefont {R.~L.}\
  \bibnamefont {Martin}}, \ and\ \bibinfo {author} {\bibfnamefont {A.~R.}\
  \bibnamefont {Bishop}},\ }\href {http://dx.doi.org/10.1038/nphys345}
  {\bibfield  {journal} {\bibinfo  {journal} {Nat Phys}\ }\textbf {\bibinfo
  {volume} {2}},\ \bibinfo {pages} {515} (\bibinfo {year} {2006})}\BibitemShut
  {NoStop}%
\bibitem [{\citenamefont {Ishioka}\ \emph {et~al.}(2008)\citenamefont
  {Ishioka}, \citenamefont {Hase}, \citenamefont {Kitajima}, \citenamefont
  {Wirtz}, \citenamefont {Rubio},\ and\ \citenamefont {Petek}}]{Ishioka2008}%
  \BibitemOpen
  \bibfield  {author} {\bibinfo {author} {\bibfnamefont {K.}~\bibnamefont
  {Ishioka}}, \bibinfo {author} {\bibfnamefont {M.}~\bibnamefont {Hase}},
  \bibinfo {author} {\bibfnamefont {M.}~\bibnamefont {Kitajima}}, \bibinfo
  {author} {\bibfnamefont {L.}~\bibnamefont {Wirtz}}, \bibinfo {author}
  {\bibfnamefont {A.}~\bibnamefont {Rubio}}, \ and\ \bibinfo {author}
  {\bibfnamefont {H.}~\bibnamefont {Petek}},\ }\href
  {http://link.aps.org/doi/10.1103/PhysRevB.77.121402} {\bibfield  {journal}
  {\bibinfo  {journal} {Phys. Rev. B}\ }\textbf {\bibinfo {volume} {77}},\
  \bibinfo {pages} {121402} (\bibinfo {year} {2008})}\BibitemShut {NoStop}%
\bibitem [{\citenamefont {Leitenstorfer}\ \emph {et~al.}(1999)\citenamefont
  {Leitenstorfer}, \citenamefont {Hunsche}, \citenamefont {Shah}, \citenamefont
  {Nuss},\ and\ \citenamefont {Knox}}]{Leitenstorfer1999}%
  \BibitemOpen
  \bibfield  {author} {\bibinfo {author} {\bibfnamefont {A.}~\bibnamefont
  {Leitenstorfer}}, \bibinfo {author} {\bibfnamefont {S.}~\bibnamefont
  {Hunsche}}, \bibinfo {author} {\bibfnamefont {J.}~\bibnamefont {Shah}},
  \bibinfo {author} {\bibfnamefont {M.~C.}\ \bibnamefont {Nuss}}, \ and\
  \bibinfo {author} {\bibfnamefont {W.~H.}\ \bibnamefont {Knox}},\ }\href
  {http://dx.doi.org/10.1063/1.123601} {\bibfield  {journal} {\bibinfo
  {journal} {Appl. Phys. Lett.}\ }\textbf {\bibinfo {volume} {74}},\ \bibinfo
  {pages} {1516} (\bibinfo {year} {1999})}\BibitemShut {NoStop}%
\bibitem [{\citenamefont {Wu}\ \emph {et~al.}(1996)\citenamefont {Wu},
  \citenamefont {Litz},\ and\ \citenamefont {Zhang}}]{Wu1996}%
  \BibitemOpen
  \bibfield  {author} {\bibinfo {author} {\bibfnamefont {Q.}~\bibnamefont
  {Wu}}, \bibinfo {author} {\bibfnamefont {M.}~\bibnamefont {Litz}}, \ and\
  \bibinfo {author} {\bibfnamefont {X.-C.}\ \bibnamefont {Zhang}},\ }\href
  {http://dx.doi.org/10.1063/1.116356} {\bibfield  {journal} {\bibinfo
  {journal} {Appl. Phys. Lett.}\ }\textbf {\bibinfo {volume} {68}},\ \bibinfo
  {pages} {2924} (\bibinfo {year} {1996})}\BibitemShut {NoStop}%
\bibitem [{\citenamefont {Hoffmann}\ \emph {et~al.}(2009)\citenamefont
  {Hoffmann}, \citenamefont {Brandt}, \citenamefont {Hwang}, \citenamefont
  {Yeh},\ and\ \citenamefont {Nelson}}]{Hoffmann2009}%
  \BibitemOpen
  \bibfield  {author} {\bibinfo {author} {\bibfnamefont {M.~C.}\ \bibnamefont
  {Hoffmann}}, \bibinfo {author} {\bibfnamefont {N.~C.}\ \bibnamefont
  {Brandt}}, \bibinfo {author} {\bibfnamefont {H.~Y.}\ \bibnamefont {Hwang}},
  \bibinfo {author} {\bibfnamefont {K.-L.}\ \bibnamefont {Yeh}}, \ and\
  \bibinfo {author} {\bibfnamefont {K.~A.}\ \bibnamefont {Nelson}},\ }\href
  {http://dx.doi.org/10.1063/1.3271520} {\bibfield  {journal} {\bibinfo
  {journal} {Appl. Phys. Lett.}\ }\textbf {\bibinfo {volume} {95}},\ \bibinfo
  {pages} {231105} (\bibinfo {year} {2009})}\BibitemShut {NoStop}%
\bibitem [{\citenamefont {Sajadi}\ \emph {et~al.}(2015)\citenamefont {Sajadi},
  \citenamefont {Wolf},\ and\ \citenamefont {Kampfrath}}]{Sajadi2015}%
  \BibitemOpen
  \bibfield  {author} {\bibinfo {author} {\bibfnamefont {M.}~\bibnamefont
  {Sajadi}}, \bibinfo {author} {\bibfnamefont {M.}~\bibnamefont {Wolf}}, \ and\
  \bibinfo {author} {\bibfnamefont {T.}~\bibnamefont {Kampfrath}},\ }\bibfield
  {booktitle} {\emph {\bibinfo {booktitle} {Optics Express}},\ }\href
  {http://www.opticsexpress.org/abstract.cfm?URI=oe-23-22-28985} {\bibfield
  {journal} {\bibinfo  {journal} {Opt. Express}\ }\textbf {\bibinfo {volume}
  {23}},\ \bibinfo {pages} {28985} (\bibinfo {year} {2015})}\BibitemShut
  {NoStop}%
\bibitem [{\citenamefont {Zhang}\ \emph {et~al.}(2009)\citenamefont {Zhang},
  \citenamefont {Zhang}, \citenamefont {Jia}, \citenamefont {Wang},\ and\
  \citenamefont {Sun}}]{Zhang2009}%
  \BibitemOpen
  \bibfield  {author} {\bibinfo {author} {\bibfnamefont {S.}~\bibnamefont
  {Zhang}}, \bibinfo {author} {\bibfnamefont {H.}~\bibnamefont {Zhang}},
  \bibinfo {author} {\bibfnamefont {T.}~\bibnamefont {Jia}}, \bibinfo {author}
  {\bibfnamefont {Z.}~\bibnamefont {Wang}}, \ and\ \bibinfo {author}
  {\bibfnamefont {Z.}~\bibnamefont {Sun}},\ }\href
  {http://link.aps.org/doi/10.1103/PhysRevA.80.043402} {\bibfield  {journal}
  {\bibinfo  {journal} {Phys. Rev. A}\ }\textbf {\bibinfo {volume} {80}},\
  \bibinfo {pages} {043402} (\bibinfo {year} {2009})}\BibitemShut {NoStop}%
\bibitem [{\citenamefont {Oomens}\ \emph {et~al.}(2006)\citenamefont {Oomens},
  \citenamefont {Sartakov}, \citenamefont {Meijer},\ and\ \citenamefont {von
  Helden}}]{Oomens2006}%
  \BibitemOpen
  \bibfield  {author} {\bibinfo {author} {\bibfnamefont {J.}~\bibnamefont
  {Oomens}}, \bibinfo {author} {\bibfnamefont {B.~G.}\ \bibnamefont
  {Sartakov}}, \bibinfo {author} {\bibfnamefont {G.}~\bibnamefont {Meijer}}, \
  and\ \bibinfo {author} {\bibfnamefont {G.}~\bibnamefont {von Helden}},\
  }\href {http://www.sciencedirect.com/science/article/pii/S1387380606002570}
  {\bibfield  {journal} {\bibinfo  {journal} {International Journal of Mass
  Spectrometry}\ }\textbf {\bibinfo {volume} {254}},\ \bibinfo {pages} {1}
  (\bibinfo {year} {2006})}\BibitemShut {NoStop}%
\bibitem [{\citenamefont {Eickemeyer}\ \emph {et~al.}(2000)\citenamefont
  {Eickemeyer}, \citenamefont {Kaindl}, \citenamefont {Woerner}, \citenamefont
  {Elsaesser},\ and\ \citenamefont {Weiner}}]{Eickemeyer2000}%
  \BibitemOpen
  \bibfield  {author} {\bibinfo {author} {\bibfnamefont {F.}~\bibnamefont
  {Eickemeyer}}, \bibinfo {author} {\bibfnamefont {R.~A.}\ \bibnamefont
  {Kaindl}}, \bibinfo {author} {\bibfnamefont {M.}~\bibnamefont {Woerner}},
  \bibinfo {author} {\bibfnamefont {T.}~\bibnamefont {Elsaesser}}, \ and\
  \bibinfo {author} {\bibfnamefont {A.~M.}\ \bibnamefont {Weiner}},\ }\href
  {http://ol.osa.org/abstract.cfm?URI=ol-25-19-1472} {\bibfield  {journal}
  {\bibinfo  {journal} {Opt. Lett.}\ }\textbf {\bibinfo {volume} {25}},\
  \bibinfo {pages} {1472} (\bibinfo {year} {2000})}\BibitemShut {NoStop}%
\bibitem [{\citenamefont {Katsuki}\ \emph {et~al.}(2013)\citenamefont
  {Katsuki}, \citenamefont {Delagnes}, \citenamefont {Hosaka}, \citenamefont
  {Ishioka}, \citenamefont {Chiba}, \citenamefont {Zijlstra}, \citenamefont
  {Garcia}, \citenamefont {Takahashi}, \citenamefont {Watanabe}, \citenamefont
  {Kitajima}, \citenamefont {Matsumoto}, \citenamefont {Nakamura},\ and\
  \citenamefont {Ohmori}}]{Katsuki2013}%
  \BibitemOpen
  \bibfield  {author} {\bibinfo {author} {\bibfnamefont {H.}~\bibnamefont
  {Katsuki}}, \bibinfo {author} {\bibfnamefont {J.~C.}\ \bibnamefont
  {Delagnes}}, \bibinfo {author} {\bibfnamefont {K.}~\bibnamefont {Hosaka}},
  \bibinfo {author} {\bibfnamefont {K.}~\bibnamefont {Ishioka}}, \bibinfo
  {author} {\bibfnamefont {H.}~\bibnamefont {Chiba}}, \bibinfo {author}
  {\bibfnamefont {E.~S.}\ \bibnamefont {Zijlstra}}, \bibinfo {author}
  {\bibfnamefont {M.~E.}\ \bibnamefont {Garcia}}, \bibinfo {author}
  {\bibfnamefont {H.}~\bibnamefont {Takahashi}}, \bibinfo {author}
  {\bibfnamefont {K.}~\bibnamefont {Watanabe}}, \bibinfo {author}
  {\bibfnamefont {M.}~\bibnamefont {Kitajima}}, \bibinfo {author}
  {\bibfnamefont {Y.}~\bibnamefont {Matsumoto}}, \bibinfo {author}
  {\bibfnamefont {K.~G.}\ \bibnamefont {Nakamura}}, \ and\ \bibinfo {author}
  {\bibfnamefont {K.}~\bibnamefont {Ohmori}},\ }\href
  {http://dx.doi.org/10.1038/ncomms3801} {\bibfield  {journal} {\bibinfo
  {journal} {Nat Commun}\ }\textbf {\bibinfo {volume} {4}} (\bibinfo {year}
  {2013})}\BibitemShut {NoStop}%
\end{thebibliography}
\end{document}


\title{\boldmath Terahertz sum-frequency excitation of a Raman-active phonon \\ (Supplemental Material)}

\author{Sebastian Maehrlein}
\email{maehrlein@fhi-berlin.mpg.de}
\affiliation{Department of Physical Chemistry, Fritz-Haber-Institut, Faradayweg 4-6, Berlin 14915, Germany}
\author{Alexander Paarmann}
\affiliation{Department of Physical Chemistry, Fritz-Haber-Institut, Faradayweg 4-6, Berlin 14915, Germany}
\author{Martin Wolf}
\affiliation{Department of Physical Chemistry, Fritz-Haber-Institut, Faradayweg 4-6, Berlin 14915, Germany}
\author{Tobias Kampfrath}
\affiliation{Department of Physical Chemistry, Fritz-Haber-Institut, Faradayweg 4-6, Berlin 14915, Germany}

\date{\today}

\maketitle

\renewcommand\thefigure{S\arabic{figure}} 

\begin{figure}[h!]
\includegraphics[width=\textwidth]{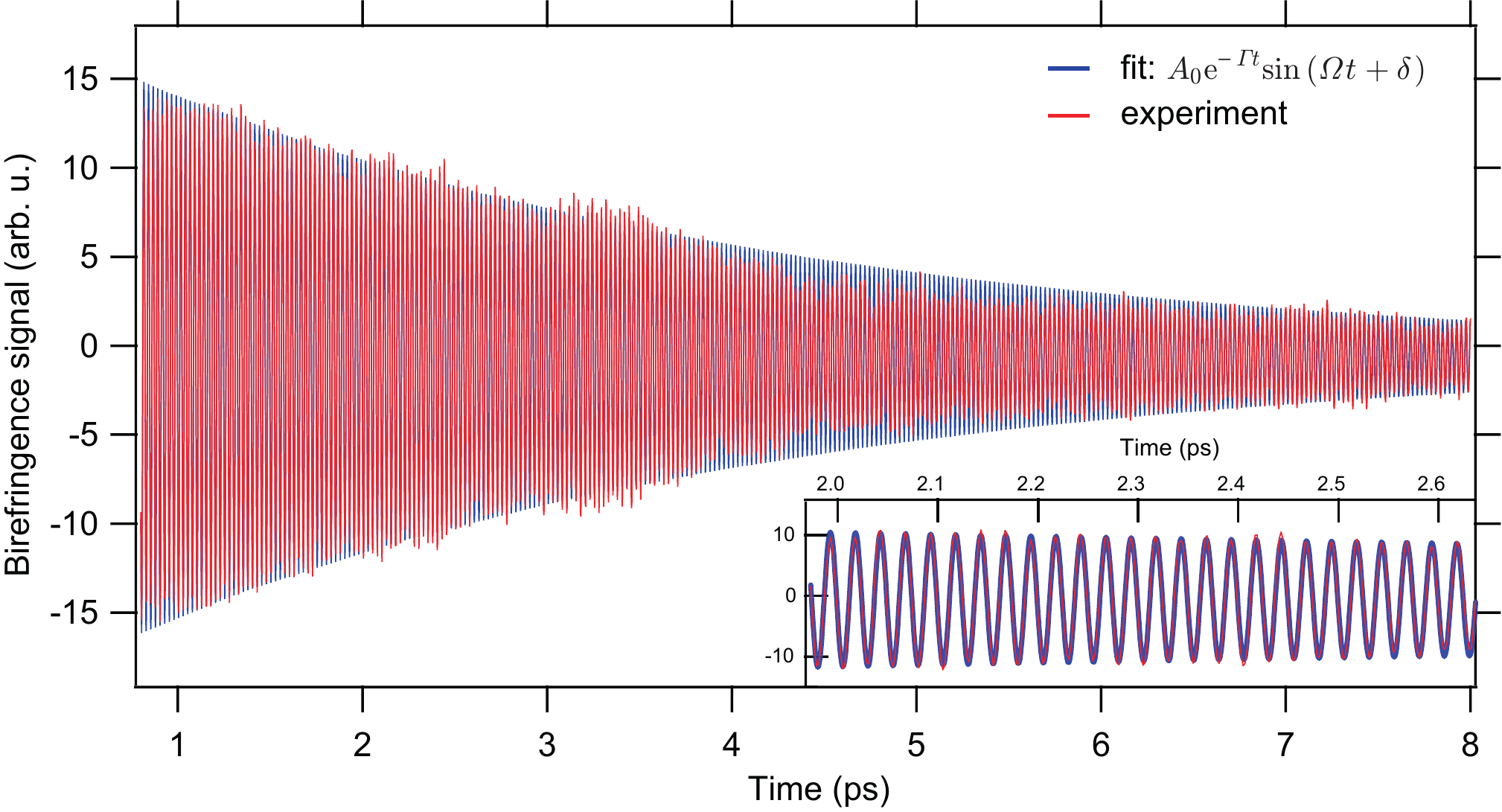}
\caption{\label{fig:FigureS1}
Time domain fit: Transient birefringence signal excited at half resonance ($\omega_0 = \Omega/2$) fitted with a single damped harmonic oscillator solution ($Q\left(t\right) = A_0 \, \mathrm{exp} \left(-\Gamma t \right) \cdot \mathrm{sin} \left( \Omega t + \delta \right)$) for $t>200$\,fs. The beating features are better reproduced by the simulation (see Fig.3(b), main paper), where the real measured electric driving field is taken into account. The inset is a zoom along the time axis to magnify the fit quality. 
}
\end{figure}

	\begin{center}
		\begin{tabular}{ l | p{4cm} | p{5cm} | p{4cm}}
			Method 	& spontaneous Raman scattering	 & Impulsive stimulated Raman scattering (ISRS)  & sum frequency excitation (SFE) \\
			\hline
			\hline
			Central frequency & $39.95 \pm 0.015$\,THz & 39.84\,THz					&				$39.95 \pm 0.01$\,THz	 \\
			Damping constant & 0.311\,ps$^{-1}$  & 0.145\,ps$^{-1}$  &  0.283\,ps$^{-1}$  							\\
			\hline
			Reference			& \cite{Solin1970} & \cite{Ishioka2006} &  \\
			\hline
			\label{tab:TableS1}
		\end{tabular}
		\end{center}
		\noindent
TABLE S1: Fit results and comparison to previous works on the $F_{2g}$-phonon mode in diamond.